\documentclass[twocolumn,conference]{IEEEtran}
\usepackage[T1]{fontenc}
\usepackage[latin9]{inputenc}
\usepackage{color}
\usepackage{amsmath}
\usepackage{amssymb}
\usepackage{graphicx}
\usepackage[unicode=true,
 bookmarks=true,bookmarksnumbered=true,bookmarksopen=true,bookmarksopenlevel=1,
 breaklinks=false,pdfborder={0 0 0},pdfborderstyle={},backref=false,colorlinks=false]
 {hyperref}
\hypersetup{pdftitle={Your Title},
 pdfauthor={Your Name},
 pdfpagelayout=OneColumn, pdfnewwindow=true, pdfstartview=XYZ, plainpages=false}
\usepackage{breakurl}

\makeatletter
\usepackage[caption=false,font=footnotesize]{subfig}
\usepackage{cite}

\makeatother

\begin{document}

\title{Performance Analysis of Low Latency Multiple Full-Duplex Selective
Decode and Forward Relays}

\author{\IEEEauthorblockN{Fatima Ezzahra Airod}\IEEEauthorblockA{Communication Systems\\
INPT\\
Rabat, Morocco\\
Email: airod@inpt.ac.ma}\and \IEEEauthorblockN{Houda Chafnaji}\IEEEauthorblockA{Communication Systems\\
INPT\\
Rabat, Morocco\\
Email: chafnaji@inpt.ac.ma}\and \IEEEauthorblockN{Halim~Yanikomeroglu\\
}\IEEEauthorblockA{Systems and Computer Engineering\\
Cartelon University \\
Ottawa, Canada\\
Email: halim@sce.carleton.ca}}

\maketitle
\thispagestyle{empty}
\begin{abstract}
In order to follow up with mission-critical applications, new features
need to be carried to satisfy a reliable communication with reduced
latency. With this regard, this paper proposes a low latency cooperative
transmission scheme, where multiple full-duplex relays, simultaneously,
assist the communication between a source node and a destination node.
First, we present the communication model of the proposed transmission
scheme. Then, we derive the outage probability closed-form for two
cases: asynchronous transmission (where all relays have different
processing delay) and synchronous transmissions (where all relays
have the same processing delay). Finally, using simulations, we confirm
the theoretical results and compare the proposed multi-relays transmission
scheme with relay selection schemes.
\end{abstract}


\section{Introduction}

Future wireless networks, i.e., 5G, open new perspectives and allow
the existence of diversified services with the aim of bringing a wide
variety of novel applications, among which we distinct mission-critical
applications. To ensure the radio communication for such applications,
very low latency as well as extreme reliability are required, whence
came, the definition of ultra-reliable and low latency communications
(URLLC). As one of flexible defined 5G service categories, URLLC needs
to be carried in cellular networks in order to enable and support
several applications, and targets important sectors namely, health,
industry and transportation. However, the requested characteristics
or functionalities will not be the same, as each application inquires
various performance requirements which makes their setting more conflicting
and challenging \cite{1,1-1}. In this context, the use of cooperation
concept provides spatial and temporal diversity, and constitutes a
good alternative to support advanced communications with increased
channel capacity \cite{2,3}. 

In general, there are various ways of relay processing in cooperative
networks, among which we distinct mainly two familiar techniques:
amplify-and-forward (AF) and decode-and-forward (DF) \cite{Laneman2}.
In AF scheme, the relay simply amplifies the received signal and forwards
it towards the destination. However, this relaying scheme suffers
from noise amplification. In the DF scheme, the relay first decodes
the signal received from the source, re-encodes and re-transmits it
to the destination. This approach suffers from error propagation when
the relay transmits an erroneously decoded data block. Selective DF,
where the relay only transmits when it can reliably decode the data
packet, has been introduced as an efficient method to reduce error
propagation \cite{Onat}. Overall, all proposed cooperative schemes
aim to increase the diversity order of the system, hence, improving
the network performance. 

Even if the full-duplex (FD) relaying mode generates loop interference
from the relay input to the relay output, it still practical to use
on cooperative relaying systems due to its spectral efficiency \cite{4,5}.
The FD relay requires the duplication of radio frequency circuits
to transmits and receives simultaneously in the same time slot and
in the same frequency band. It has been shown that the FD mode still
feasible even with the presence of significant loop interference \cite{4},
especially with recent advances noted in antenna technology and signal
processing techniques. In \cite{6}, a novel technique for self-interference
cancellation using antenna cancellation was depicted for FD transmissions.
In the same context, through passive suppression and active self-interference
cancellation mechanisms, an experiment study was proposed in \cite{7}.
Hence, these practical growths incite authors to adopt FD communications
in their research, thus, get rid of spectral inefficiency caused by
half-duplex (HD) relaying mode. 

In cooperative systems, one or multiple relays may be used to assist
transmission between a source and a destination nodes. The application
of the relay selection principle on FD system permits the merging
of space diversity as well as the spectral efficiency \cite{10}.
Therefore, several works in the literature have considered the relay
selection concept applied to their studied multiple relays systems
\cite{10,11,12}. The best proved relay selection policy for FD cooperative
networks is the optimal relay selection (OS) \cite{10,12}. This scheme
takes into consideration the global channel state information (CSI)
of the source to relay channels as well as that of the relay to destination
channels. So, despite its proved performance, the OS induces more
system overhead \cite{10,13,14}, hence, more system latency. With
the aim of reducing the system latency and the implementation complexity,
partial relay selection (PS) scheme that requires just the CSI knowledge
of one hop, were introduced in \cite{10}. To the best of our knowledge,
only few works carried the multiple relays model without relay selection.
In \cite{8}, the performance of HD multiple decode-and-forward system,
were investigated for non identical distributed channels. Recently,
FD-AF cooperative system were studied \cite{2}. The authors proposed
a forced delayed FD relaying scheme, where an iterative successive
interference cancellation model was used to withdraw the accumulation
effect between signals at the destination. In this paper, we propose
a multiple FD relaying scheme, where non-controlled selective decode
and forward (SDF) relays, simultaneously, assist the communication
between a source and destination nodes. First, we derive the outage
probability closed-form of the proposed system. Then, as a benchmark,
we investigate the performances comparison with the OS and the PS
relay selection schemes. 

The rest of the paper is organized as follows: Section$\,$\ref{sec:Communication-model}
presents the communication model of the proposed transmission scheme.
The outage probability of multiple FD-SDF relays is derived in Section$\,$\ref{sec:Outage-probability}.
In Section$\,$\ref{sec:Results-and-discussion}, Numerical results
are shown and discussed. The paper is concluded in Section$\,$\ref{sec:Conclusion}. 

$\vphantom{}$

\noindent \textcolor{black}{\emph{$\qquad$Notations}}
\begin{itemize}
\item $x$, $\mathbf{x}$, and $\mathbf{X}$ denote, respectively, a scalar
quantity, a column vector, and a matrix.
\item $\mathcal{CN}\left(\mu,\sigma\right)$ represents a circularly symmetric
complex Gaussian distribution with mean $\mu$ and variance $\sigma$.
\item $\delta_{m,n}$ is the Kronecker symbol, i.e., $\delta_{m,n}=1$ for
$m=n$ and $\delta_{m,n}=0$ for $m\neq n$.
\item $\left(.\right)^{\star}$\textcolor{black}{,$\left(.\right)^{\top}$,
and $\left(.\right)^{\mathrm{H}}$ are conjugate, the transpose, and
the }Hermitian transpose, respectively\textcolor{black}{. }
\item $\mathbb{C}$ is set of complex number.
\item \textcolor{black}{For $\mathbf{x}\in\mathbb{C}^{N\times1}$, $\mathbf{x}_{f}$
denotes the discrete Fourier transform (DFT) of $\mathbf{x}$, i.e.,
$\mathbf{x}_{f}=\mathbf{U}_{N}\mathbf{x}$, with $\mathbf{U}_{N}$
is a unitary $N\times N$ matrix whose $\left(m,n\right)$th element
is $\left(\mathbf{U}_{N}\right)_{m,n}=\frac{1}{\sqrt{N}}e^{-j(2\pi mn/N)}$,
$j=\sqrt{-1}$.}
\item \textcolor{black}{$\left|.\right|$ denotes the }absolute value.
\item $\mathbb{{E}}\left\{ .\right\} $ is used to denote the statistical
expectation.
\item $\mathrm{Pr}\left(X\right)$ is the probability of occurrence of the
event $X$.
\end{itemize}

\section{\label{sec:Communication-model}Communication Model}

\begin{figure}[tbh]
\includegraphics[scale=0.55]{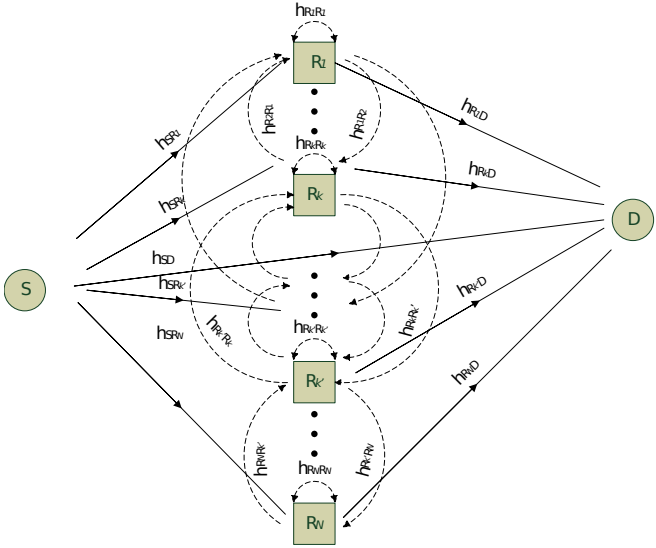}

\caption{\label{fig:The-FD-SDF_model}The FD SDF multi-relay system.}
\end{figure}

We consider a multi-relay cooperative system, where a set $\mathcal{R}$
of $N$ FD-relays $(\mathrm{R_{\mathit{k}}})$, $(k=1,...,N)$ assists
the communication between a source $(\mathrm{S})$ and a destination
$(\mathrm{D})$, as depicted in Fig. \ref{fig:The-FD-SDF_model}.
Since all relays operate in FD mode, we take into account the residual
self-interference (RSI) generated from relay's input to relay's output,
as well as inter-relay interference (IRI). 

The source-destination $\mathrm{S\rightarrow D}$, source-relay $\mathrm{S\rightarrow R_{\mathit{k}}}$,
the relay interference $\mathrm{R_{\mathit{k'}}\rightarrow R_{\mathit{k}}}$,
i.e., RSI $(k=k')$ and IRI $(k\neq k')$, and relay-destination $\mathrm{R_{\mathit{k}}\rightarrow D}$
channels, are represented by $h_{\mathrm{\mathrm{ab}}},$ with $\mathrm{ab}\,\epsilon\left\{ \mathrm{SD},\,\mathrm{SR_{\mathit{k}}},\,\mathrm{R_{\mathit{k'}}R_{\mathit{k}}},\,\mathrm{R_{\mathit{k}}D}\right\} $.
In this paper, all channels are assumed independent identically distributed
(i.i.d.) zero mean circularly symmetric complex Gaussian $\sim\mathcal{CN}(0,\sigma_{\mathrm{ab}}^{2})$.\textcolor{red}{{}
}We assume a perfect CSI at the receiver nodes and limited CSI at
the transmitter nodes, i.e., the transmitter is only aware of the
processing delay at the relay nodes.

In this work, we consider all relays are operating using SDF relaying
mode, where the relay transmits only when it can correctly decode
the source message. The received signals, at time instance $i$, at
relay $\mathrm{R_{\mathit{k}}}$ and destination $\mathrm{D}$ are,
respectively, given by

\begin{equation}
\begin{aligned}y_{\mathrm{\mathrm{R_{\mathit{k}}}}}(i) & =\sqrt{P_{\textrm{S}}}h_{\mathrm{SR_{\mathit{k}}}}x_{\mathrm{s}}(i)+\sum_{\mathrm{R}_{\mathit{\mathit{k'}}}\in\mathcal{R}_{L}}\sqrt{P_{\mathrm{R}}}h_{\mathrm{R_{\mathit{k'}}R_{\mathit{k}}}}x_{\mathrm{s}}(i-\tau_{\mathit{k'}})\\
 & +n_{\mathrm{R_{\mathit{k}}}}(i),\\
 & =\sqrt{P_{\textrm{S}}}h_{\mathrm{SR_{\mathit{k}}}}x_{\mathrm{s}}(i)+\underset{\mathrm{RSI+IRI}}{\underbrace{V_{\mathrm{R}_{\mathit{k}}}(i)}}+n_{\mathrm{R}_{\mathit{k}}}(i),
\end{aligned}
\end{equation}

\begin{equation}
\begin{aligned}y_{\mathrm{D}}(i) & =\underset{\mathrm{Direct\,+\,Relayed\,signal}}{\underbrace{\sqrt{P_{\textrm{S}}}h_{\mathrm{SD}}x_{\mathrm{s}}(i)+\sum_{\mathrm{R}_{\mathit{\mathit{k}}}\in\mathcal{R}_{L}}\sqrt{P_{\mathrm{R}}}h_{\mathrm{\mathrm{R}_{\mathit{k}}D}}x_{\mathrm{s}}(i-\tau_{\mathit{k}})}}\\
 & +\underset{\mathrm{Noise}}{\underbrace{n_{\mathrm{D}}(i)}},
\end{aligned}
\label{eq:Dest_receiv_sig}
\end{equation}
where $P_{\textrm{S}}$ and $P_{\mathrm{R}}$ denote, respectively,
the transmit power of $\mathrm{S}$ and $\mathrm{R}_{\mathit{\mathit{k}}}$,
$x_{\mathrm{s}}(i)$ is the source transmitted signal at channel use
$i$ with $\mathbb{{E}}\left[x_{\mathrm{s}}(i)x_{\mathrm{s}}^{\star}(i')\right]=\delta_{i,i'}$,
and $\mathcal{R}_{\mathit{L}}\subset\mathcal{R}$ denotes the set
of $L$ relays that correctly decode the source message. $n_{\mathrm{R_{\mathit{k}}}}\sim\mathcal{CN}(0,N_{\mathrm{R}})$
and $n_{\mathrm{D}}\sim\mathcal{CN}(0,N_{\mathrm{D}})$ respectively
denote, a zero-mean complex additive white Gaussian noise at the relay
$\mathrm{R}_{\mathit{\mathit{k}}}$ and the destination $\mathrm{D}$.
Without loss of generality and for the sake of presentation, we assume
$N_{\mathrm{D}}=N_{\mathrm{R}}=1$. The processing delay at relay
$\mathrm{R}_{\mathit{\mathit{k}}}$ is denoted $\tau_{k}$ , $V_{\mathrm{R}_{\mathit{\mathit{k}}}}(i)$
covers the RSI$+$IRI at a relay $\mathrm{R}_{\mathit{\mathit{k}}}$
after undergoing all known cancellation techniques and practical isolation
\cite{5,15}. $V_{\mathrm{R}_{\mathit{\mathit{k}}}}(i)$ is assumed
to be equivalent to a zero mean complex Gaussian random variable $\sim\mathcal{CN}(0,\sigma_{\mathrm{RSI,R_{\mathit{k}}}}^{2}+\sigma_{\mathrm{IRI,R_{\mathit{k}}}}^{2})$,
with $\sigma_{\mathrm{RSI,\mathit{k}}}^{2}=\sigma_{\mathrm{R_{\mathit{k}}R_{\mathit{k}}}}^{2}$
and $\sigma_{\mathrm{IRI,\mathit{k}}}^{2}={\displaystyle \sum_{\underset{k'\neq k}{\mathrm{R}_{\mathit{\mathit{k'}}}\in\mathcal{R}_{L}}}}\sigma_{\mathrm{R_{\mathit{k'}}R_{\mathit{k}}}}^{2}$.

From (\ref{eq:Dest_receiv_sig}), we can see that the destination
node will receive the source node transmitted signal $x_{\mathrm{s}}$
at different time instance due to the processing delay $\tau_{k}$
at the relay $\mathrm{R}_{\mathit{k}}$. In order to alleviate the
inter-symbol interference (ISI) caused by the delayed signal, equalization
is needed at the destination side. For that purpose, we propose a
cyclic-prefix (CP) transmission at the source side in order to perform
frequency-domain equalization (FDE) at the destination node. 

In this paper, we assume that all channel gains change independently
from one block to another and remain constant during one block of
$T+\tau_{\mathrm{CP}}$ channel uses, where $T$ represents the number
of transmitted code-words and $\tau_{\mathrm{CP}}$ the CP length
($\tau_{\mathrm{CP}}\geq\mathit{\underset{k}{\mathrm{max}}\mathrm{(\tau_{\mathit{k}})}}$).
Hence, (\ref{eq:Dest_receiv_sig}) can be rewritten in vector form
to jointly take into account the $T+\tau_{\mathrm{CP}}$ received
signal as \cite{key-7}

\begin{equation}
\mathbf{y}_{D}=\boldsymbol{{\mathcal{H}}}\mathbf{x}_{\mathrm{s}}+\mathbf{n},
\end{equation}
where $\mathbf{y}_{\mathrm{D}}=\left[y_{\mathrm{D}}\left(0\right),...,y_{\mathrm{D}}\left(T-1\right)\right]^{\top}\in\mathbb{C}^{T\times1}$,
$\mathbf{x}_{\mathrm{s}}=\left[x_{\mathrm{s}}\left(0\right),...,x_{\mathrm{s}}\left(T-1\right)\right]^{\top}\in\mathbb{C}^{T\times1}$,
with $\mathbf{n}=\left[n_{\mathrm{D}}\left(0\right),...,n_{\mathrm{D}}\left(T-1\right)\right]^{\top}\in\mathbb{C}^{T\times1}$
and $\boldsymbol{{\mathcal{H}}}\in\mathbb{C}^{T\times T}$ is a circulant
matrix that can be decomposed as 

\begin{equation}
\boldsymbol{{\mathcal{H}}}=\mathbf{U}_{T}^{H}\boldsymbol{{\Lambda}}\mathbf{U}_{T},
\end{equation}
where $\boldsymbol{{\Lambda}}$ is a diagonal matrix whose $\left(i,i\right)$-th
element is

\begin{equation}
\lambda_{i}=\sqrt{P_{\textrm{S}}}h_{\mathrm{SD}}+{\displaystyle \sum_{\mathrm{R}_{\mathit{\mathit{k}}}\in\mathcal{R}_{L}}}\sqrt{P_{\mathrm{R}}}h_{\mathrm{R_{\mathit{k}}D}}\textrm{e}^{-j\left(2\pi i\frac{\tau_{k}}{T}\right)}.\label{eq:lambda}
\end{equation}
The signal $\mathbf{y}_{\mathrm{D}}$ can be therefore represented
in the frequency domain as

\begin{equation}
\mathbf{y}_{\mathrm{D}_{f}}=\boldsymbol{{\Lambda}}\mathbf{x}_{\mathrm{s}_{f}}+\mathbf{n}_{f}.
\end{equation}

At the destination, the instantaneous end-to-end equivalent signal-to-interference
and noise ratio (SINR), at frequency bin $i$, is expressed as

\begin{equation}
\begin{aligned}\gamma_{i}= & \lambda_{i}\lambda_{i}^{H}\\
= & P_{\textrm{S}}|h_{\mathrm{SD}}|^{2}+\alpha_{L}+A,
\end{aligned}
\end{equation}
where $\alpha_{L}=P_{\mathrm{R}}|{\displaystyle \sum_{\mathrm{R}_{\mathit{\mathit{k}}}\in\mathcal{R}_{L}}}h_{\mathrm{R_{\mathit{k}}D}}e^{-j\left(2\pi i\frac{\tau_{k}}{T}\right)}|^{2}$,
$A=2\sqrt{P_{\textrm{S}}}\sqrt{P_{\mathrm{R}}}{\displaystyle \sum_{\mathrm{R}_{\mathit{\mathit{k}}}\in\mathcal{R}_{L}}}\left(|h_{\mathrm{SD}}h_{\mathrm{R_{\mathit{k}}D}}^{\text{*}}|\cos\left(2\pi i\frac{\tau_{k}}{T}+\theta_{\mathit{k}}\right)\right)$
and $\theta_{\mathit{k}}=\textrm{angle}\left(h_{\mathrm{SD}}h_{\mathrm{R_{\mathit{k}}D}}^{\text{*}}\right)$.

\section{\label{sec:Outage-probability}Outage Probability}

In this section, we derive the proposed transmission scheme outage
probability. For that purpose, let's first introduce the instantaneous
SINRs for each link. The received instantaneous SINR of $\mathrm{S}\rightarrow\mathrm{D}$,
$\mathrm{S}\rightarrow\mathrm{R_{\mathit{k}}}$ and $\mathrm{R_{\mathit{k}}}\rightarrow\mathrm{D}$
links are, respectively, denoted $\gamma_{\mathrm{SD}}=P_{\textrm{S}}|h_{\mathrm{SD}}|^{2}$,
$\gamma_{\mathrm{R_{\mathit{\mathit{k}}}D}}=P_{\mathrm{R}}|h_{\mathrm{R_{\mathit{k}}D}}|^{2}${\footnotesize{}
}and $\gamma_{\mathrm{SR_{\mathit{k}}}}=\frac{P_{\textrm{S}}|h_{\mathrm{\mathrm{SR_{\mathit{k}}}}}|^{2}}{P_{\mathrm{R}}\left(\sigma_{\mathrm{RSI,R_{\mathit{k}}}}^{2}+\sigma_{\mathrm{IRI,R_{\mathit{k}}}}^{2}\right)+1}$.
Note that all SINRs are exponentially distributed random variables.

The multiple SDF FD relay system outage probability can be expressed
as

\begin{equation}
\begin{aligned}\mathcal{P}_{out} & =P_{out}^{\mathrm{S\rightarrow D}}\underset{\mathrm{R^{'}}\in\mathit{\mathcal{R}}}{\prod}P_{out}^{\mathrm{S\rightarrow R^{'}}}\\
 & +\sum_{L=1}^{N}\mathit{\sum_{\mathcal{R}_{L}}P_{out}^{\mathrm{S}\mathcal{R_{\mathit{L}}}\mathrm{D}}}\underset{\mathrm{R}\in\mathcal{R}_{L}}{\prod}\left(1-P_{out}^{\mathrm{S\rightarrow R}}\right)\underset{\mathrm{R^{'}}\in\overline{\mathcal{R}}_{L}}{\prod}P_{out}^{\mathrm{S\rightarrow R^{'}}},
\end{aligned}
\label{eq:sdfout}
\end{equation}
where $\mathcal{R}_{L}$ denotes the set of $L$ relays not in outage
and $\overline{\mathcal{R}}_{L}\triangleq\mathit{\mathcal{R}}\setminus\mathit{\mathcal{R}}_{L}$.
$P_{out}^{\mathrm{S\rightarrow D}}$ and $P_{out}^{\mathrm{S\rightarrow R}}$
denote respectively, the outage probability of $\mathrm{S\rightarrow D}$
link and $\mathrm{S\rightarrow R}$ link, and can be expressed as
\begin{eqnarray}
P_{out}^{\mathrm{S\rightarrow D}}= & \mathrm{Pr}(\gamma_{\mathrm{SD}}<\eta)= & 1-\textrm{e}^{-\frac{\eta}{P_{\textrm{S}}\sigma_{\mathrm{SD}}^{2}}}\nonumber \\
P_{out}^{\mathrm{S\rightarrow R_{\mathit{k}}}}= & \mathrm{Pr}(\gamma_{\mathrm{SR_{\mathit{k}}}}<\eta)= & 1-\textrm{e}^{-\frac{\eta\left(P_{\mathrm{R}}\left(\sigma_{\mathrm{RSI,R_{\mathit{k}}}}^{2}+\sigma_{\mathrm{IRI,R_{\mathit{k}}}}^{2}\right)+1\right)}{P_{\textrm{S}}\sigma_{\mathrm{SR}}^{2}}},\label{eq:SR_SD}
\end{eqnarray}
where $\eta=2^{r\left(\frac{T+\tau_{CP}}{T}\right)}-1$, with $r$
is the bit rate per channel use. Note that the factor $\frac{T+\tau_{CP}}{T}$
means that the transmission of $T$ useful code-words occupies $T+\tau_{CP}$
channel uses. $P_{out}^{\mathrm{S}\mathcal{R_{\mathit{L}}}\mathrm{D}}$
denotes, the outage probability of a cooperative system where a set
$\mathcal{R}_{L}$ of $L$ relays assist the communication between
node $\mathrm{S}$ and node $\mathrm{D}$, and it can be derived as
follows:

\begin{equation}
P_{out}^{\mathrm{S}\mathcal{R_{\mathit{L}}}\mathrm{D}}=\mathrm{Pr}\left(\frac{1}{T+\tau_{CP}}\sum_{i=0}^{T-1}\log_{2}(1+\gamma_{i})<r\right).\label{eq:14}
\end{equation}

To derive the closed form expression of (\ref{eq:14}), we consider
two cases, i.e., the asynchronous transmission $(\tau_{\mathit{k}}\neq\tau_{\mathit{k'}},\,\forall\mathit{k\neq k'})$
and the synchronous transmission $(\tau_{\mathit{k}}=\tau_{\mathit{k'}}=\tau,\,\forall\mathit{k\neq k'})$. 
\begin{itemize}
\item \textbf{Asynchronous transmission}
\end{itemize}
In the asynchronous transmission, all relays forward signals to the
destination with different delay processing, i.e., $(\tau_{\mathit{k}}\neq\tau_{\mathit{k'}},\,\forall\mathit{k\neq k'})$.
Inspired from \cite{16}, we have ${\displaystyle \sum_{i=0}^{T-1}}\log_{2}(1+\gamma_{i})=\sum_{i=0}^{T-1}\log_{2}\left\{ \left(1+P_{\textrm{S}}|h_{\mathrm{SD}}|^{2}+\alpha_{L}\right)\times\left(1+\frac{A}{1+P_{\textrm{S}}|h_{\mathrm{SD}}|^{2}+\alpha_{L}}\right)\right\} $,
and thereby, we get,
\begin{align}
\sum_{i=0}^{T-1}\log_{2}(1+\gamma_{i}) & =\sum_{i=0}^{T-1}\log_{2}\left(1+P_{\textrm{S}}|h_{\mathrm{SD}}|^{2}+\alpha_{L}\right)\nonumber \\
 & +\sum_{i=0}^{T-1}\log_{2}\left(1+\frac{A}{1+P_{\textrm{S}}|h_{\mathrm{SD}}|^{2}+\alpha_{L}}\right).\label{eq:approx1}
\end{align}
Thanks to arithmetic-geometric mean inequality for complex number,
we get $P_{\textrm{S}}|h_{\mathrm{SD}}|^{2}+\alpha_{L}>A$. Thus,
using the first Taylor expansion, $\log_{2}\left(1+\frac{A}{1+P_{\textrm{S}}|h_{\mathrm{SD}}|^{2}+\alpha_{L}}\right)\approx\frac{1}{\ln\left(2\right)}\frac{A}{1+P_{\textrm{S}}|h_{\mathrm{SD}}|^{2}+\alpha_{L}}$.
Noting that {\small{}${\displaystyle \sum_{i=0}^{T-1}}\cos\left(2\pi i\frac{\tau_{k}}{T}+\theta_{k}\right)=0$}.
Therefore, the second term in (\ref{eq:approx1}){\scriptsize{} }vanishes{\scriptsize{}.}
Thus, (\ref{eq:approx1}) can be approximated as

{\footnotesize{}
\begin{align}
\sum_{i=0}^{T-1}\log_{2}(1+\gamma_{i}) & \approx\sum_{i=0}^{T-1}\log_{2}\left(1+P_{\textrm{S}}|h_{\mathrm{SD}}|^{2}+\alpha_{L}\right)\nonumber \\
 & =\sum_{i=0}^{T-1}\log_{2}\left(1+P_{\textrm{S}}|h_{\mathrm{SD}}|^{2}+P_{\mathrm{R}}|h_{\mathrm{R_{\mathit{L}}D}}|^{2}+\alpha_{L-1}\right.\nonumber \\
 & +\left.\beta_{L}\right)\nonumber \\
 & =\sum_{i=0}^{T-1}\log_{2}\left(1+P_{\textrm{S}}|h_{\mathrm{SD}}|^{2}+P_{\mathrm{R}}|h_{\mathrm{R_{\mathit{L}}D}}|^{2}+\alpha_{L-1}\right)+\nonumber \\
 & \sum_{i=0}^{T-1}\log_{2}\left(1+\frac{\beta_{L}}{1+P_{\textrm{S}}|h_{\mathrm{SD}}|^{2}+P_{\mathrm{R}}|h_{\mathrm{R_{\mathit{L}}D}}|^{2}+\alpha_{L-1}}\right),\label{eq:secondapp}
\end{align}
}with $\beta_{L}=2P_{\mathrm{R}}{\displaystyle \sum_{\mathrm{R}_{\mathit{\mathit{k}}}\in\mathcal{R}_{L}}}\left(|h_{\mathrm{R_{\mathit{L}}D}}h_{\mathrm{R_{\mathit{k}}D}}^{\text{*}}|\cos\left(2\pi i\frac{\tau_{L}-\tau_{k}}{T}+\varphi_{L,k}\right)\right)$
and $\varphi_{L,k}=\textrm{angle}\left(h_{\mathrm{R_{\mathit{L}}D}}h_{\mathrm{R_{\mathit{k}}D}}^{\text{*}}\right)$.
Noting that $1+P_{\textrm{S}}|h_{\mathrm{SD}}|^{2}+P_{\mathrm{R}}|h_{\mathrm{R_{\mathit{L}}D}}|^{2}+\alpha_{L-1}>P_{\mathrm{R}}|h_{\mathrm{R_{\mathit{L}}D}}|^{2}+\alpha_{L-1}\geq\beta_{L}$
and using the same mathematical manipulations as before, we can easily
proof that the second term in (\ref{eq:secondapp}) vanishes. Repeating
the same mathematical manipulations, we found that (\ref{eq:secondapp})
can be approximated as
\begin{equation}
\sum_{i=0}^{T-1}\log_{2}(1+\gamma_{i})\approx T\log_{2}\left(1+P_{\textrm{S}}|h_{\mathrm{SD}}|^{2}+P_{\mathrm{R}}\sum_{\mathrm{R}_{\mathit{\mathit{k}}}\in\mathcal{R}_{L}}|h_{\mathrm{R_{\mathit{k}}D}}|^{2}\right).\label{eq:async_approx-1}
\end{equation}

From (\ref{eq:async_approx-1}), we can see that using equalization
at the destination side, for asynchronous transmission, allows to
virtually separate different spatial paths and thereby achieve a full
spatial diversity. Therefore, $P_{out}^{\mathrm{S}\mathcal{R_{\mathit{L}}}\mathrm{D}}$
can be derived as
\begin{align}
P_{out}^{S\mathcal{R_{\mathit{L}}}\mathrm{D}}= & \mathrm{Pr}\left(\frac{T}{T+\tau_{CP}}\log_{2}\left(1+\gamma_{\mathrm{SD}}+\sum_{\mathrm{R}_{\mathit{\mathit{k}}}\in\mathcal{R}_{L}}\gamma_{\mathrm{R_{\mathit{\mathit{k}}}D}}\right)<r\right)\nonumber \\
= & \mathrm{Pr}\left(\gamma_{\mathrm{SD}}+\sum_{\mathrm{R}_{\mathit{\mathit{k}}}\in\mathcal{R}_{L}}\gamma_{\mathrm{R_{\mathit{\mathit{k}}}D}}<\eta\right)\nonumber \\
= & \intop_{0}^{\eta}\mathrm{Pr}\left(\gamma_{\mathrm{SD}}<\eta-y\right)\times f_{{\displaystyle \sum_{\mathrm{R}_{\mathit{\mathit{k}}}\in\mathcal{R}_{L}}}\gamma_{\mathrm{R_{\mathit{\mathit{k}}}D}}}(y)\textrm{d}y.\label{eq:}
\end{align}

For simplicity, we consider all \textcolor{black}{relays experience
the same}\textcolor{red}{{} }$\mathrm{R_{\mathit{k}}}\rightarrow\mathrm{D}$
link\textcolor{red}{{} }\textcolor{black}{quality, i.e., }$\sigma_{\mathrm{RD}}^{2}=\sigma_{\mathrm{R_{\mathit{k}}D}}^{2},\:\forall k$\textcolor{black}{.
Therefore, }${\displaystyle \sum_{\mathrm{R}_{\mathit{\mathit{k}}}\in\mathcal{R}_{L}}}\gamma_{_{\gamma_{\mathrm{R_{\mathit{\mathit{k}}}D}}}}$
follows gamma distribution with parameters $L$ and $\overline{\gamma}_{\mathrm{RD}}=P_{\mathrm{R}}\sigma_{\mathrm{RD}}^{2}$,
and with probability distribution function (pdf) $f_{{\displaystyle \sum_{\mathrm{R}_{\mathit{\mathit{k}}}\in\mathcal{R}_{L}}}\gamma_{_{\gamma_{\mathrm{R_{\mathit{\mathit{k}}}D}}}}}(x)=\frac{1}{\overline{\gamma}_{\mathrm{RD}}}\textrm{e}^{-\frac{x}{\overline{\gamma}_{\mathrm{RD}}}}\frac{\left(\frac{x}{\overline{\gamma}_{\mathrm{RD}}}\right)^{L-1}}{\left(L-1\right)!}$\textcolor{black}{.
So according}ly, after some manipulations, we get the expression of
$P_{out}^{\mathrm{S}\mathcal{R_{\mathit{L}}}\mathrm{D}}$ as depicted
below:
\begin{align}
P_{out}^{S\mathcal{R_{\mathit{L}}}D}= & \frac{\gamma\left(L,\frac{\eta}{\overline{\gamma}_{\mathrm{RD}}}\right)}{\Gamma(L)}-\frac{\textrm{e}^{-\frac{\eta}{\overline{\gamma}_{\mathrm{SD}}}}}{\Gamma(L)}\times\nonumber \\
 & \left(\frac{\overline{\gamma}_{\mathrm{SD}}}{\overline{\gamma}_{\mathrm{SD}}-\overline{\gamma}_{\mathrm{RD}}}\right)^{L}\times\gamma\left(L,\eta\frac{\overline{\gamma}_{\mathrm{SD}}-\overline{\gamma}_{\mathrm{RD}}}{\overline{\gamma}_{\mathrm{RD}}\overline{\gamma}_{\mathrm{SD}}}\right),\label{eq:srd}
\end{align}
where $\overline{\gamma}_{\mathrm{SD}}=P_{\textrm{S}}\sigma_{\mathrm{SD}}^{2}$,
$\Gamma(L)=\left(L-1\right)!$ is the factorial of $L-1$, and $\gamma(n,x)$
presents the lower incomplete Gamma function which is given by $\intop_{0}^{x}t^{n-1}\textrm{e}^{-t}\textrm{d}t$
\cite[8.350.1]{17}. Thereby, by substituting (\ref{eq:SR_SD}) and
(\ref{eq:srd}) into (\ref{eq:sdfout}), we get the closed form expression
of the outage probability for the asynchronous case.
\begin{itemize}
\item \textbf{Synchronous transmission}
\end{itemize}
In the synchronous transmission, all relays forward signals to the
destination with the same delay processing. Therefore, $\lambda_{i}$
in (\ref{eq:lambda}) can be expressed as $\lambda_{i}=\sqrt{P_{\textrm{S}}}h_{\mathrm{SD}}+\sqrt{P_{\mathrm{R}}}\left({\displaystyle \sum_{\mathrm{R}_{\mathit{\mathit{k}}}\in\mathcal{R}_{L}}}h_{\mathrm{R_{\mathit{k}}D}}\right)\textrm{e}^{-j\left(2\pi i\frac{\tau_{k}}{T}\right)}$.
We see clearly that the synchronous transmission is equivalent to
one relay system with $\mathrm{R\rightarrow D}$ channel $h_{\mathrm{syn}}={\displaystyle \sum_{\mathrm{R}_{\mathit{\mathit{k}}}\in\mathcal{R}_{L}}}h_{\mathrm{R_{\mathit{k}}D}}\sim\mathcal{CN}(0,{\displaystyle \sum_{\mathrm{R}_{\mathit{\mathit{k}}}\in\mathcal{R}_{L}}}\sigma_{\mathrm{R}_{k}\mathrm{D}}^{2})$
and received instantaneous SINR $\gamma_{\mathrm{syn}}=P_{\mathrm{R}}|h_{\mathrm{syn}}|^{2}$.
Thus, synchronous transmission represents the worst scenario where
adding more relays does not add any diversity to the system \cite{2}. 

By referring to the proof in \cite{16}, $P_{out}^{\mathrm{S}\mathcal{R_{\mathit{L}}}\mathrm{D}}$
can be derived as

\begin{equation}
\begin{aligned}P_{out}^{\mathrm{S}\mathcal{R_{\mathit{L}}}\mathrm{D}}\approx & \mathrm{Pr}\left(\frac{T}{T+\tau_{\mathrm{CP}}}\log_{2}(1+\gamma_{\mathrm{SD}}+\gamma_{\mathrm{syn}})<r\right)\\
= & \mathrm{Pr}\left(\gamma_{\mathrm{SD}}+\gamma_{\mathrm{syn}}<\eta\right)\\
= & \intop_{0}^{\eta}\mathrm{Pr}\left(\gamma_{\mathrm{SD}}<\eta-y\right)\times f_{\gamma_{\mathrm{syn}}}(y)\textrm{d}y
\end{aligned}
\label{eq:12synch}
\end{equation}
where $f_{\gamma_{\mathrm{syn}}}(y)=\frac{1}{\overline{\gamma}_{\mathrm{\mathrm{syn}}}}\textrm{e}^{-\frac{y}{\overline{\gamma}_{\mathrm{\mathrm{syn}}}}}$
represents the pdf of {\small{}$\gamma_{\mathrm{syn}}$, with }$\overline{\gamma}_{\mathrm{\mathrm{syn}}}=P_{\mathrm{R}}{\displaystyle \sum_{\mathrm{R}_{\mathit{\mathit{k}}}\in\mathcal{R}_{L}}}\sigma_{\mathrm{RD}}^{2}${\small{}.
Hence, the (\ref{eq:12synch}) can be expressed as}{\small \par}

\begin{equation}
\begin{aligned}P_{out}^{\mathrm{S}\mathcal{R_{\mathit{L}}}\mathrm{D}}= & \left(1-\textrm{e}^{-\frac{\eta}{\overline{\gamma}_{\mathrm{\mathrm{syn}}}}}\right)-\\
 & \left(\frac{\overline{\gamma}_{\mathrm{SD}}}{\overline{\gamma}_{\mathrm{SD}}-\overline{\gamma}_{\mathrm{\mathrm{syn}}}}\right)\textrm{e}^{-\bar{\gamma}_{SD}\eta}\times\left(1-\textrm{e}^{-\eta\left(\frac{\overline{\gamma}_{\mathrm{SD}}-\overline{\gamma}_{\mathrm{\mathrm{\mathrm{syn}}}}}{\overline{\gamma}_{\mathrm{\mathrm{\mathrm{syn}}}}\overline{\gamma}_{\mathrm{SD}}}\right)}\right).
\end{aligned}
\label{eq:srdsyn}
\end{equation}

Finally, by substituting (\ref{eq:SR_SD}) and (\ref{eq:srdsyn})
into (\ref{eq:sdfout}), we get the closed form expression of synchronous
case outage probability.

\section{\label{sec:Results-and-discussion}Numerical Results }

In this section, using Monte-Carlo simulations, we evaluate the performance
of the studied FD Multi-relay system, with non controlled SDF relays.
For comparison, we consider two relay selection schemes, i.e., the
OS as the high latency relay selection scheme and the PS as the low
latency scheme. Note that both considered relay selection schemes
require more system overhead than the proposed scheme, and hence,
more system latency. \textcolor{black}{For simplicity, we assume all
relays experience the same channel quality, i.e.,} $\sigma_{\mathrm{SR}}^{2}=\sigma_{\mathrm{SR_{\mathit{k}}}}^{2}$,
$\sigma_{\mathrm{RD}}^{2}=\sigma_{\mathrm{R_{\mathit{k}}D}}^{2}$,
$\sigma_{\mathrm{RSI}}^{2}=\sigma_{\mathrm{RSI,\mathit{k}}}^{2}$,
and $\sigma_{\mathrm{IRI}}^{2}=\sigma_{\mathrm{IRI,\mathit{k}}}^{2}$,
$\forall k=1,...,N.$ Besides, for all simulations, we assume that
$\sigma_{\mathrm{SD}}^{2}=0\mathrm{\,dB}$, $r=\mathrm{2bps/Hz}$,
$\mathrm{\mathit{T}=500,}$ and $\mathit{\tau}_{\mathrm{CP}}=10$.
For a fair comparison, we set the relay transmit power of the proposed
multi-relay scheme to $P_{\mathrm{R}}=\frac{E_{\mathrm{R}}}{L}$ and
the relay selection schemes to $E_{\mathrm{R}}$. 

\begin{figure}[tbh]
\includegraphics[scale=0.62]{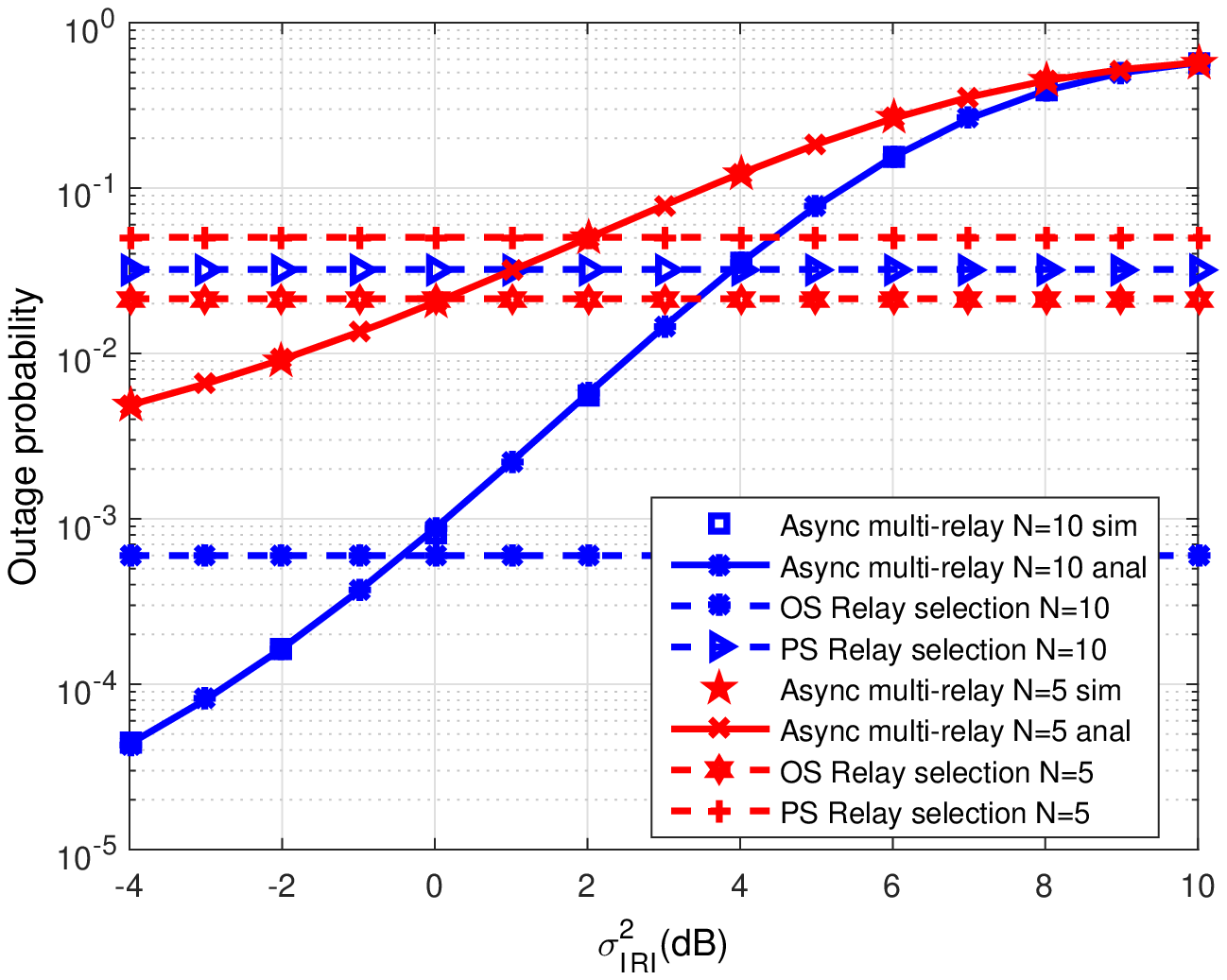}

\caption{\label{fig:Outage-probability-versusIRI}Outage probability versus
the IRI of asynchronous case for $\sigma_{\mathit{\mathrm{SR}}}^{2}=8\,\mathrm{dB}$,
$\sigma_{\mathrm{RD}}^{2}=10\,\mathrm{dB}$, $\sigma_{\mathrm{RSI}}^{2}=0\mathrm{\,dB}$,
and $P_{\textrm{S}}=E_{\textrm{R}}=5\,\mathrm{dB}.$}

\includegraphics[scale=0.62]{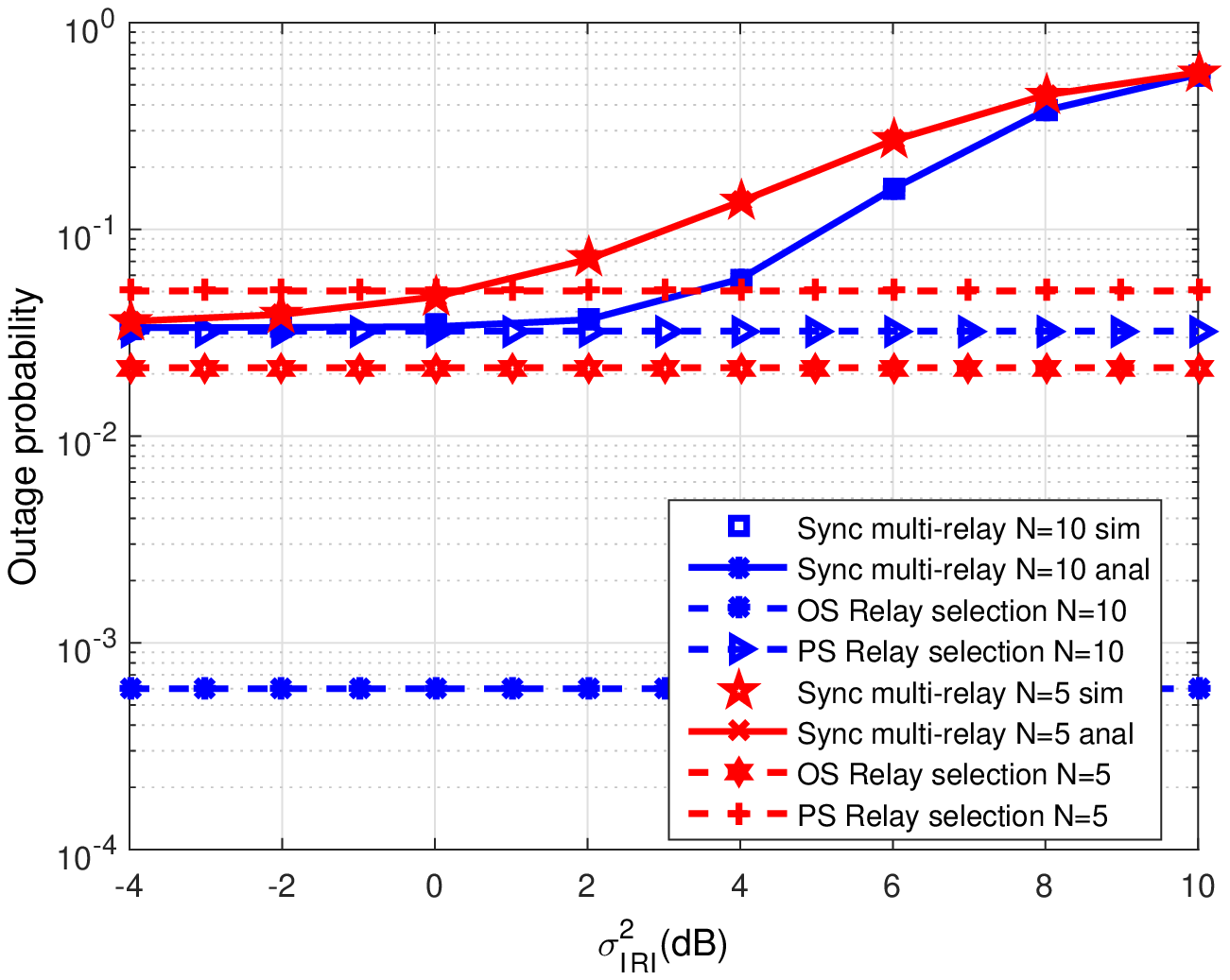}

\caption{\label{fig:Outage-probability-versusIRI-1}Outage probability versus
the IRI of synchronous case for $\sigma_{\mathrm{SR}}^{2}=8\mathrm{\,dB}$,
$\sigma_{\mathrm{RD}}^{2}=10\,\mathrm{dB}$, $\sigma_{\mathrm{RSI}}^{2}=0\mathrm{\,dB}$,
and $P_{\textrm{S}}=E_{\textrm{R}}=5\,\mathrm{dB}.$}
\end{figure}

Fig. \ref{fig:Outage-probability-versusIRI} and Fig. \ref{fig:Outage-probability-versusIRI-1}
illustrate the performances of the investigated system model in section$\,$\ref{sec:Communication-model}.
They represent, respectively, asynchronous and synchronous cases,
where the outage probability of the three relaying schemes, cited
above, are plotted versus $\sigma_{\mathrm{IRI}}^{2}$. Moreover,
to point out the impact of the number of relays on the system performances,
the evaluation is performed for two different number of relays, i.e.,
$N=5$ and $N=10$, for a fixed value of RSI, i.e., $\sigma_{\mathrm{RSI}}^{2}=0\mathrm{dB}$.
First, we notice that the simulation results match perfectly with
the theoretical analysis, obtained in section$\,$\ref{sec:Outage-probability},
for both synchronous and asynchronous cases. From Fig. \ref{fig:Outage-probability-versusIRI},
that represents the best scenario where all relays are asynchronous,
we can see clearly that the system performances become better as $N$
increases, mainly due to the additional spatial diversity. Furthermore,
depending on the inter-relay-interference level at the relays, i.e.,
$\sigma_{\mathrm{IRI}}^{2}$, the three considered relaying schemes
outperform each other. In term of outage probability, when the system
suffers from high IRI, OS scheme offers the best performance gain
but at the price of high system overhead. For low IRI, i.e.,$\sigma_{\mathrm{IRI}}^{2}<\sigma_{\mathrm{RSI}}^{2}$,
the proposed multi-relay scheme becomes the best choice in term of
both outage probability and latency. Note that, due to the distance
between the transmit and receive antennas that reduces naturally the
IRI, \textcolor{black}{we should conside}r $\sigma_{\mathrm{IRI}}^{2}<\sigma_{\mathrm{RSI}}^{2}$
for practical scenarios. Now, we turn to the worst scenario where
all relays are synchronous. From Fig. \ref{fig:Outage-probability-versusIRI-1},
we notice that the curves of synchronous case have a very bad slope
and saturate at low $\sigma_{\mathrm{IRI}}^{2}$. In fact, in the
synchronous case adding more relays does not add any spatial diversity
to the system. Even for a such bad scenario, we can see, from Fig.
\ref{fig:Outage-probability-versusIRI-1}, that for $N=5$, the multi-relay
transmission scheme outperforms the moderate latency relay selection
PS at low $\sigma_{\mathrm{IRI}}^{2}$.

\begin{figure}[tbh]
\includegraphics[scale=0.65]{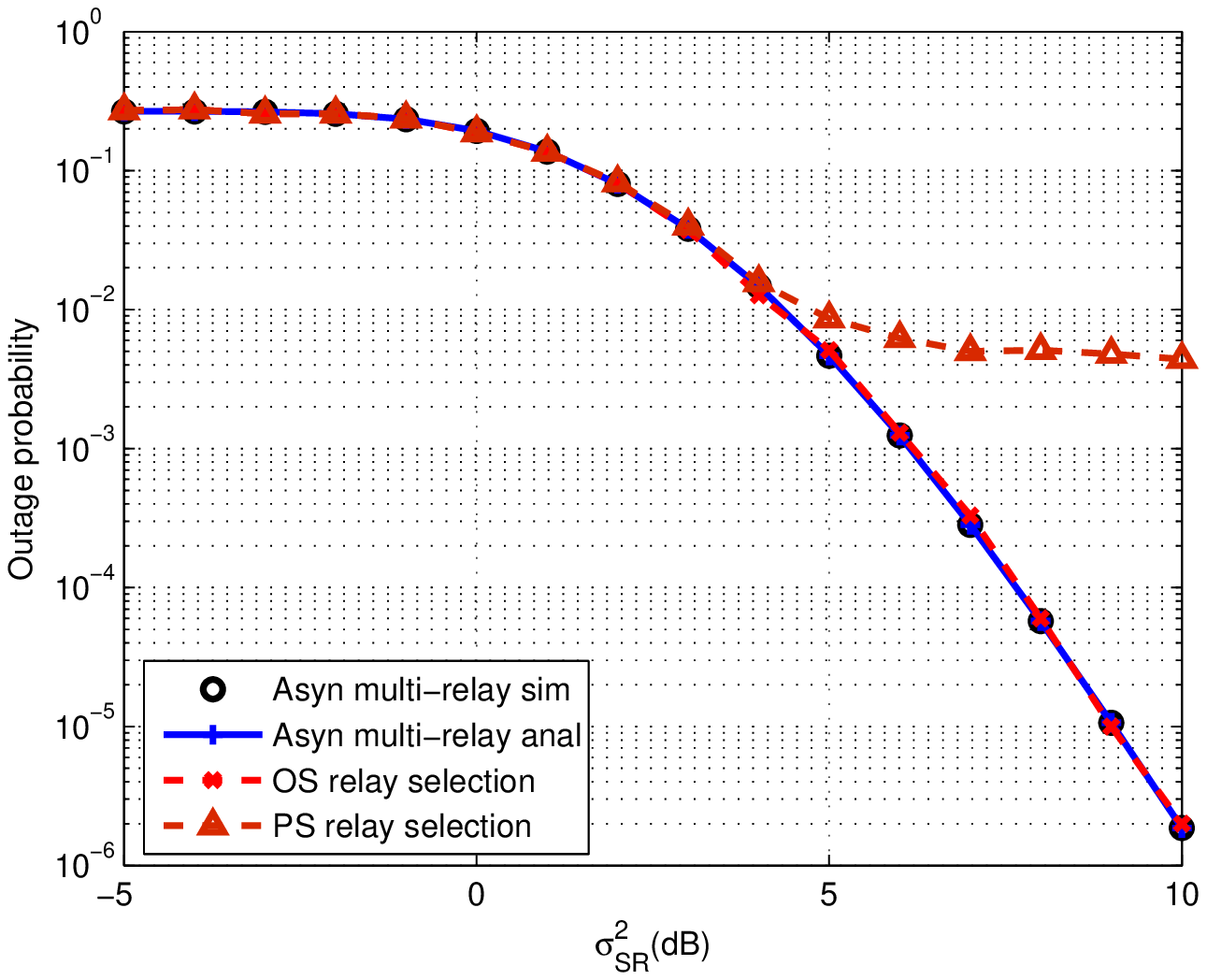}

\caption{\label{fig:Outage-probability-versusSRHighRD}Asynchronous outage
probability versus $\sigma_{\mathrm{SR}}^{2}$ for $N=10$, $\sigma_{\mathrm{RD}}^{2}=10\,\mathrm{dB}$,
$\sigma_{\mathrm{RSI}}^{2}=\sigma_{\mathrm{IRI}}^{2}=0\mathrm{\,dB}$,
and $P_{\textrm{S}}=E_{\textrm{R}}=10\,\mathrm{dB}.$}
\end{figure}

\begin{figure}[tbh]
\includegraphics[scale=0.65]{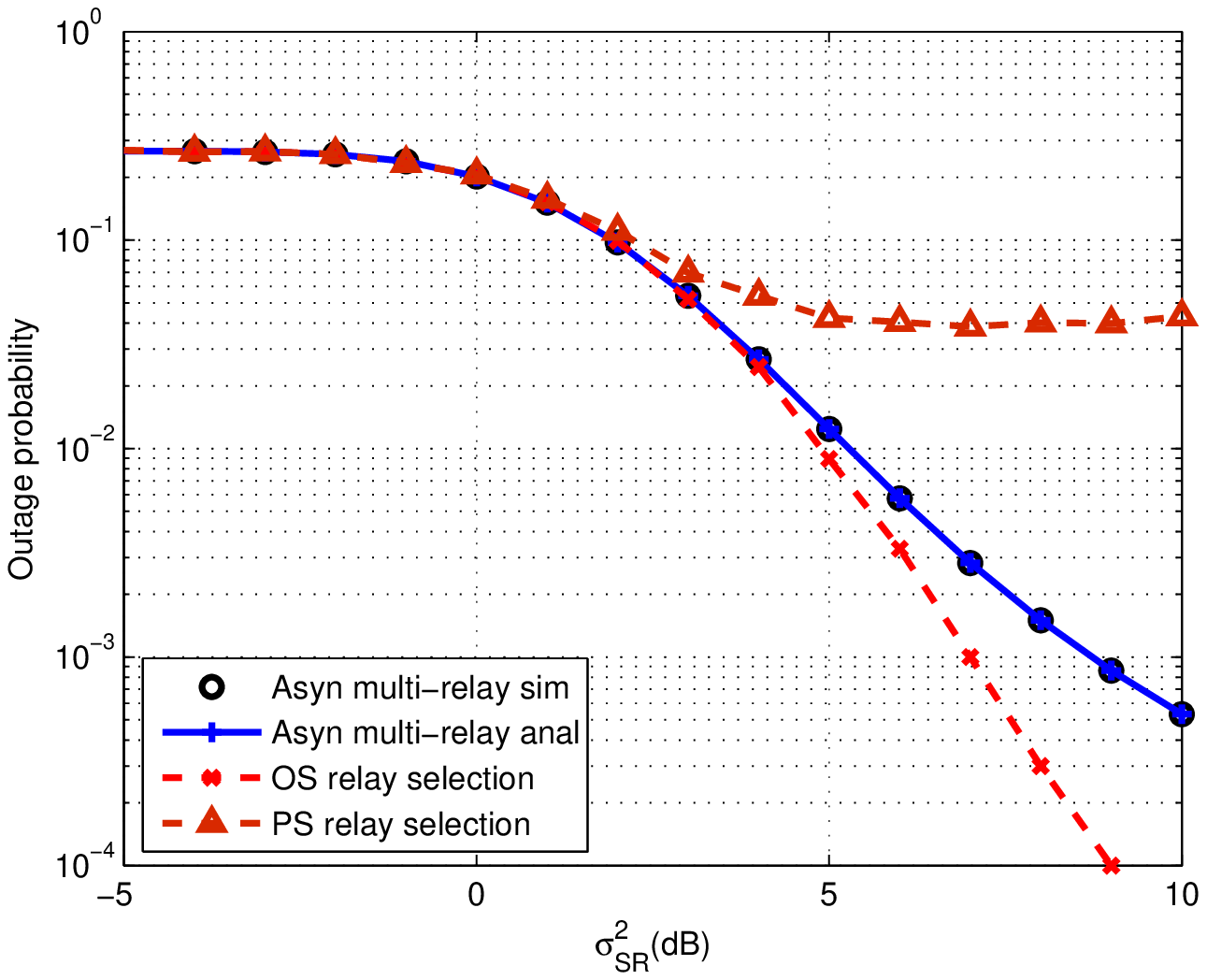}

\caption{\label{fig:Outage-probability-versusSRLowRD}Asynchronous outage probability
versus $\sigma_{\mathrm{SR}}^{2}$ for $N=10$, $\sigma_{\mathrm{RD}}^{2}=0\mathrm{\,dB}$,
$\sigma_{\mathrm{RSI}}^{2}=\sigma_{\mathrm{IRI}}^{2}=0\mathrm{\,dB}$
and $P_{\textrm{S}}=E_{\textrm{R}}=10\,\mathrm{dB}.$}
\end{figure}

Now, we focus on the asynchronous scenario and evaluate the outage
probability of the studied system versus $\sigma_{\mathrm{SR}}^{2}$.
In Fig. \ref{fig:Outage-probability-versusSRHighRD}, we consider
the scenario of a strong $\mathrm{R_{\mathit{k}}}\rightarrow\mathrm{D}$
link, i.e., $\sigma_{\mathrm{RD}}^{2}=10\,\mathrm{dB}$, and we can
see clearly that the proposed multi-relay system and the OS scheme
offer the same performances, while outperforming the PS scheme with
the increase of $\sigma_{\mathrm{SR}}^{2}$. In Fig. \ref{fig:Outage-probability-versusSRLowRD},
as the $\mathrm{R_{\mathit{k}}}\rightarrow\mathrm{D}$ link quality
decreases, i.e., $\sigma_{\mathrm{RD}}^{2}=0\,\mathrm{dB}$, we start
to notice that the OS scheme, provides better performances than the
multi-relay system when $\sigma_{\mathrm{SR}}^{2}\geq4\mathrm{\,dB}$.
This is due to the fact that, in OS scheme, the relaying transmit
power $E_{\mathrm{R}}$ is fully used by the best $\mathrm{R_{\mathit{k}}}\rightarrow\mathrm{D}$
link while, in multi-relay scheme, the relaying transmit power $E_{\mathrm{R}}$
is\textcolor{red}{{} }\textcolor{black}{shared }equally between $L$
relay links, i.e., $P_{\mathrm{R}}=\frac{E_{\mathrm{R}}}{L}$. Even
though, the proposed scheme still performs better than the PS scheme. 

\section{\label{sec:Conclusion}Conclusion}

In this paper, we proposed a low latency cooperative transmission
scheme, where multiple FD-SDF relays, simultaneously, assist the communication
between a source node and a destination node. First, the analytical
expression of the outage probability were derived for two cases, i.e.,
asynchronous and synchronous transmissions. Then, using Monte-carlo
simulations, we compared the proposed multi-relays transmission scheme
with two different relay selection schemes, i.e., the OS scheme requiring
the knowledge of global CSI and the PS scheme requiring the knowledge
of partial CSI. Simulation results reveal that the proposed multi-relay
transmission scheme and relay selection schemes outperform each other
in term of outage probability, depending on IRI, number of relays,
and channel links quality. As the proposed multiple FD cooperative
relaying scheme does not require any central component, thus, getting
rid of relay selection signaling messages and thereby, reducing the
system latency while increasing the system diversity, we can say that
it can be considered as a good candidate for very low latency applications.


\begin{thebibliography}{10}
\bibitem{1}R. Abreu, P. Mogensen, and K. I. Pedersen, ``Pre-scheduled
resources for retransmissions in ultra-reliable and low latency communications'',
in \textit{Proc. IEEE WCNC}, San Francisco, USA, March 2017.

\bibitem{1-1}H. Shariatmadari, S. Iraji, Z. Li, M. A. Uusitalo, and
R. Jäntti, ``Optimized transmission and resource allocation strategies
for ultra-reliable communications'', in \textit{Proc. IEEE PIMRC},
Valencia, Spain, September 2016.

\bibitem{2}J. Han, J. Baek, S. Jeon, and J. Seo, ``Cooperative networks
with amplify-and-forward multiple-full-duplex relays'', \textit{IEEE
Transactions on Wireless Communications}, Vol. 13, no. 4, pp. 2137
- 2149, April 2014.

\bibitem{3}A. F. M. Shahen Shah and Md. Shariful Islam, ``A survey
on cooperative communication in wireless networks'', \textit{I.J.
Intelligent Systems and Applications}, pp. 66-78, June 2014.

\bibitem{Laneman2}J. N. Laneman, D. Tse, and G. W. Wornell, ``Cooperative
diversity in wireless networks: Efficient protocols and outage behavior'',
\textit{IEEE Trans. Inform. Theory}, vol. 50, no. 12, pp. 3062-3080,
December. 2004.

\bibitem{Onat}F. Atay Onat, H. Yanikomeroglu, and S. Periyalwar,
``Relay-assisted spatial multiplexing in wireless fixed relay networks'',
\textit{IEEE GLOBECOM}, San Francisco, USA, Nov.- Dec. 2006.

\bibitem{4}T. Riihonen, S. Werner, R. Wichman, and E. Zacarias, \textquotedbl{}On
the feasibility of full-duplex relaying in the presence of loop interference\textquotedbl{},
in \textit{Proc. IEEE SPAWC}, Perugia, Italy, June 2009.

\bibitem{5}T. Riihonen, S. Werner, and R. Wichman, \textquotedblleft Optimized
gain control for single-frequency relaying with loop interference\textquotedblright ,
\textit{IEEE Trans Wireless Commun}, vol. 8, no. 6, pp. 2801\textendash 2806,
June 2009.

\bibitem{6}J. I. Choi, M. Jain, K. Srinivasan, P. Levis, and S. Katti,
\textquotedblleft Achieving single channel, full duplex wireless communication,\textquotedblright{}
in  \textit{Proc. ACM MobiCom}, Chicago, Illinois, USA, September
2010.

\bibitem{7}M. Duarte, C. Dick, and A. Sabharwal, ``Experiment-driven
characterization of full-duplex wireless systems,'' \textit{IEEE
Transactions on Wireless Communications}, vol. 11, no. 12, pp. 4296-4307,
May 2012.

\bibitem{10}I. Krikidis, H. A. Suraweera, P. J. Smith, and C. Yuen,
``Full-duplex relay selection for amplify-and-forward cooperative
networks'', \textit{IEEE Transactions on Wireless Communications},
vol. 11, no. 12, pp. 4381-4393, December 2012.

\bibitem{11}Y. Wang, Y. Xu, N. Li, W. Xie, K. Xu, and X. Xia, ``Relay
selection of full-duplex decode-and-forward relaying over Nakagami-m
fading channels'', \textit{IET Communications}, vol. 10, no. 12,
pp. 170-179, 2016.

\bibitem{12}S. S. Ikki and M. H. Ahmed, ``Performance analysis of
adaptive decode-and-forward cooperative diversity networks with best-relay
selection'', \textit{IEEE Transactions on Communications}, vol. 58,
no. 1, January 2010.

\bibitem{13}Z. Ding, I. Krikidis, B. Sharif, and H. V. Poor, ``Wireless
information and power transfer in cooperative networks with spatially
random relays'', \textit{IEEE Transactions on Wireless Communications},
vol. 13, no. 8, pp. 4440-4453, 2014.

\bibitem{14}A. Bletsas, A. Khisti, D. P. Reed, and A. Lippman, ``A
simple cooperative diversity method based on network path selection'',
\textit{IEEE Journal on Selected Areas in Communications}, vol. 24,
no. 3, pp. 659-672, March 2006.

\bibitem{8}N. C. Beaulieu and J. Hu, `` A closed-form expression
for the outage probability of decode-and-forward relaying in dissimilar
Rayleigh fading channels'', \textit{IEEE Communications Letters},
vol. 10, no. 12, 2006.

\bibitem{15}M. Jain, J. Choi, T. Kim, D. Bharadia, S. Seth, K. Srinivasan,
P. Levis, S. Katti, and, P. Sinha, \textquotedblleft Practical, real-time,
full duplex wireless\textquotedblright , in \textit{Proc. ACM} \textit{MobiCom},
Las Vegas, Nevada, USA, September 2011. 

\bibitem{key-7}H. Chafnaji, T. Ait-Idir, H. Yanikomeroglu, and S.
Saoudi, ``Turbo packet combining for relaying schemes over multiantenna
broadband channels'', \textit{IEEE Transactions on Vehicular Technology},
vol. 61, no. 7, pp. 2965-2977, 2012.

\bibitem{16}M. G. Khafagy, A. Ismail, M. S. Alouini, and S. Aïssa,
``On the outage performance of full-duplex selective decode-and-forward
relaying,'' \textit{IEEE Communications Letters}, vol. 17, no. 6,
pp. 1180-1183, April 2013.

\bibitem{17}I. Gradshteyn and I. Ryzhik,\textit{ }``Table of Integrals,
Series and Products'', $\textrm{\ensuremath{7^{th}}}$ edition. London:
Academic Press, 2007.
\end{thebibliography}
\end{document}